\documentclass[
    ,final            
  ]
  {aipproc}

\layoutstyle{8x11double}

\begin{document}

\title{CCD Photometry of the Pleiades Delta Scuti Star  V650 Tauri}


\author{L. Fox Machado}{
 address={Observatorio Astron\'omico Nacional,
Instituto de Astronom\'{\i}a, Universidad Nacional Aut\'onoma de M\'exico,
A.P. 877, Ensenada, BC 22860, Mexico}
}

\author{R. Michel}{
  address={Observatorio Astron\'omico Nacional, Instituto de Astronom\'{\i}a, Universidad Nacional Aut\'onoma de M\'exico, A.P. 877, Ensenada, BC 22860, Mexico}
}

\author{M. Alvarez}{
  address={Observatorio Astron\'omico Nacional, Instituto de Astronom\'{\i}a, Universidad Nacional Aut\'onoma de M\'exico, A.P. 877, Ensenada, BC 22860, Mexico}
}

\author{C. Zurita}{
  address={Instituto de Astrof\'{\i}sica de Canarias, C/V\'{\i}a L\'actea s/n, E-38205, La Laguna, Tenerife, Spain}
}

\author{J.N. Fu}{
  address={Department of Astronomy, Beijing Normal University, Beijing 100875, China}
}

\begin{abstract}
The  preliminary results of a three-site CCD photometric campaign
are reported. The  $\delta$ Scuti variable V650 Tauri  belonging to
the Pleiades cluster was observed photometrically for 14 days on
three continents during 2008 November.  An overall run of 164 hr of
data was collected. At least five significant frequencies for V650
Tauri have been detected.

\end{abstract}

\maketitle


\section{Introduction}

$\delta$ Scuti variables are stars with masses between 1.5 and 2.5
$M_{\odot}$ located at the intersection of the classical Cepheid
instability strip with the main sequence. These variables are
thought to be excellent laboratories for probing the internal
structure of intermediate mass stars. Intents of modelling $\delta$
Scuti stars belonging to open clusters have been performed recently
(e.g. [1], [2], [3]). Although the constraints imposed by the
cluster parameters have proved to be very useful when modelling an
ensemble of $\delta$ Scuti stars, more detected frequencies in
individual stars would improve current seismic studies.

\bigskip
The target star V650 Tau (HD 23643, $V=7^{\rm m}.79$, A7)  was
identified as a short-period pulsating variable by Breger (1972).
Intensive observations performed by the STEPHI network in November
1990, revealed four frequency peaks in V650 Tau [4]. One-site CCD
photometric observations carried out by [5] in November-December
1993, confirmed the results obtained by the STEPHI campaign. Since
then, no new observations of V650 Tauri have been performed.

\bigskip
The present paper provides preliminary observational results of a
three-site campaign on V650 Tauri in 2008.

\section{Observations and data reduction}

Three observatories were involved in the observational campaign.
They are listed in Table~\ref{tab:tel} together with the telescopes
and instruments used. Table~\ref{tab:log} gives the log of
observations. A total amount of 164 hours of useful data were
obtained  from the three sites.

\begin{figure}[]
 \centering
 \includegraphics[width=7cm]{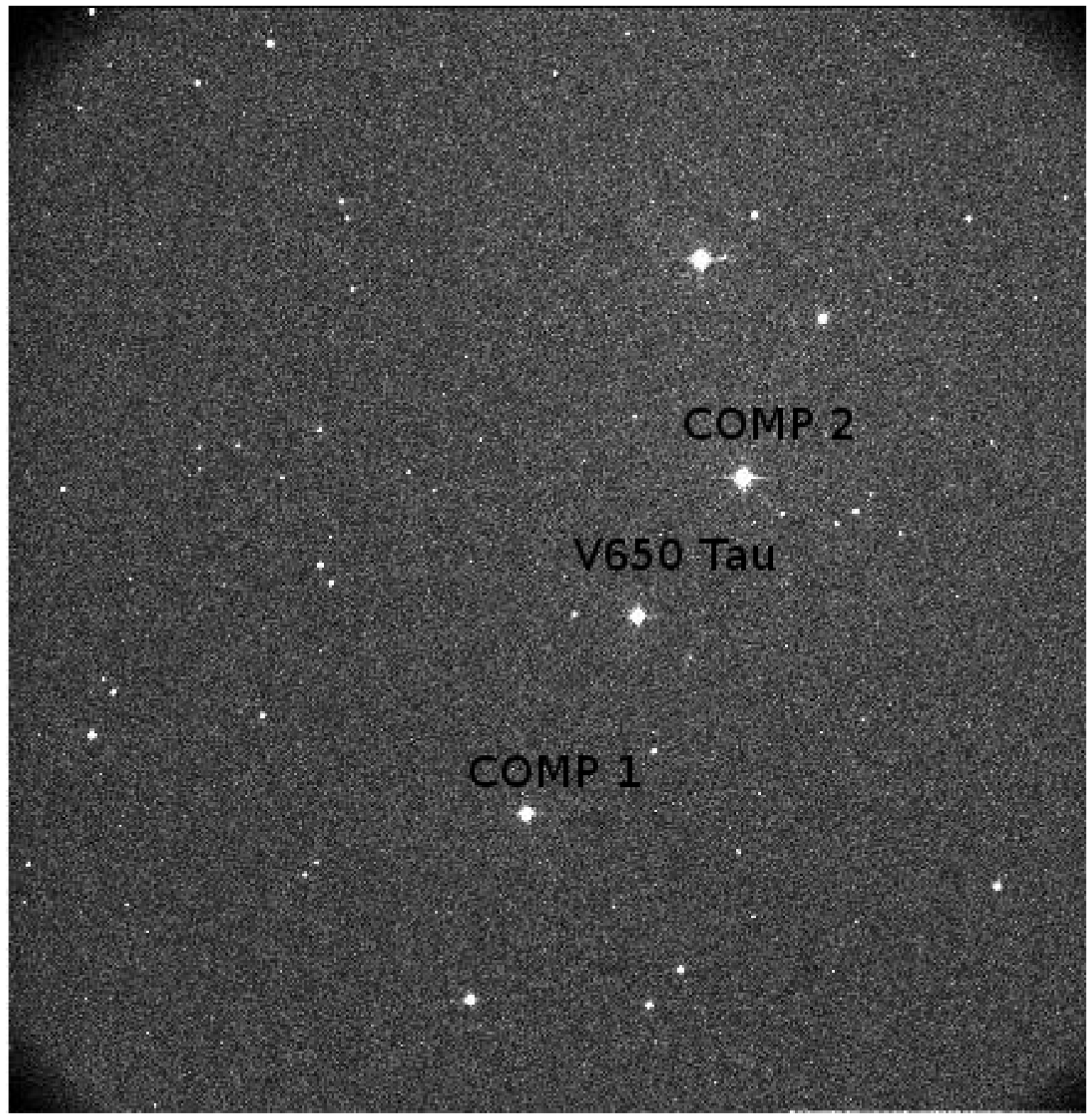}
 \caption{Image of the CCD field-of-view ($20' \times 20'$) of the Xing Long Observatory.
 The positions of the target and comparison stars are indicated in the figure. North is up and East is left.}
 \label{fig:field}
\end{figure}

\medskip
The observations were obtained through a Johnson $V$ filter except
at the SPM observatory where a Str\"omgren $y$ filter was used.
Table~\ref{tab:stars} shows the main observational parameters
corresponding to the target and comparison stars as taken from the
SIMBAD database operated by CDS (Centre de Donn\'ees astronomique de
Strasbourg). Two comparison stars have been used during the
observations depending on the constraints set by the field of view
of the CCD's and sizes of the telescopes. The first one, HD 23605
($V=6^{\rm m}.99$, F5), is a suitable comparison star considering
its brightness and spectral type. However, this star could not be
observed neither at Teide nor at San Pedro Martir observatory
because it is so bright that at these telescopes the  CCD detectors
saturated in a few seconds of exposure time. Rather at these sites
we observed the comparison star, HD 23653 ($V=7^{\rm m}.71$, K0),
since its magnitude is similar to that of the target star.
Figure~\ref{fig:field} shows a typical image of the CCD's field of
view ($20' \times 20'$) at the 0.50m telescope of the Xing Long
observatory.

\medskip
 Sky flats,
dark and bias exposures were taken every night at all sites. All
data were calibrated and reduced using standard IRAF routines.
Aperture photometry was implemented to extract the instrumental
magnitudes of the stars. The differential magnitudes were normalized
by subtracting the mean of differential magnitudes for each night.
In Figure~\ref{fig:curves} the entire light curves V650 Tau - Comp 2
are presented.

\begin{table}[!t]\centering
 \setlength{\tabcolsep}{1.0\tabcolsep}
 \caption{List of instruments and telescopes involved in the campaign. Observer's abbreviations
 correspond to initial of the co-authors.} \label{tab:tel}

  \begin{tabular}{lccc}
\hline
Observatory&  Telescope& Instrument & Observers  \\
\hline
 Observatorio del Teide (OT, Spain) &0.80m & 2048x2048 CCD & CZ \\
 Observatorio San Pedro M\'artir (SPM, Mexico) &0.84m& 1024x1024 CCD&LFM, RM, MA \\
 Xing Long Station (XL, China) &0.50m& 1024x1024 CCD&JNF\\
\hline
\end{tabular}
\end{table}

\begin{table}
\caption{Log of observations. Observing time is expressed in hours.}
\begin{tabular}{ccclll}

\hline

   Day   &    Date 2008  &   HJD    & SPM & XL           & OT    \\
         &               &2454774+ &      &              &       \\
 \hline
    01    &       Nov 11  & 07   &  0.70  &    &   5.70   \\
    02    &       Nov 12  & 08    & 3.58  &    &   8.80   \\
    03    &       Nov 13  & 09    & 4.76  &    &   -   \\
    04    &       Nov 14  & 10    & 9.68  &    &   1.08   \\
    05    &       Nov 15  & 11    & 9.42  &    &   -   \\
    06    &       Nov 16  & 12    & 9.79  &  -  &   8.17  \\
    07    &       Nov 17  & 13    & 7.64  &  -  &   -   \\
    08    &       Nov 18  & 14    & 10.15 & 10.28&    -   \\
    09    &       Nov 19  & 15    & 10.26 & 10.19&   -   \\
   10    &       Nov 20  & 16    & 10.33  & 10.59&    -   \\
    11   &       Nov 21  & 17    &  7.22  & 10.36&   -   \\
    12   &       Nov 22  & 18    &  -    &  10.66&   -   \\
    13   &       Nov 23  & 19    & -     &  -   &       -  \\
    14   &       Nov 24  & 20    &  -  &  4.98 &   -    \\
    \hline
  Total&  observing &  time & SPM &   XL &  OT   \\
  Nov 11&  Nov 24  & 164.34 & 83.52 &57.06 &  23.75 \\
\hline
\end{tabular}
\label{tab:log}
\end{table}

\begin{table}[ht]
  \caption{Observational properties of the stars observed in the
campaign.}
  \begin{tabular}{lccccccc}
  \hline
  Star        & HD    &  ST  & $V$ &  $B-V$  & $U-B$ & $v \sin i$ & $\beta$ \\
              &       &      &        &      &   & $(\mathrm{km\, s}^{-1})$  \\
  \hline
  V650 Tau     & 23643 & A7  &  7.79 & $+$0.25 & $+$0.14 &  219 & 2.823  \\
  Comparison 1 & 23605 & F5  &  6.99 & $+$0.50 & $+$0.09 &  - & 2.653  \\
  Comparison 2 & 23653 & K0  &  7.71 & $+$1.27 & $+$1.12 & - & -  \\
  \hline
  \end{tabular}
  \label{tab:stars}
\end{table}

\section{Spectral analysis}

 The period analysis has been performed by means of
standard Fourier analysis and least-squares fitting. In particular,
the amplitude spectra of the differential time series were obtained
by means of Period04 package [6], which considers Fourier as well as
multiple least-squares algorithms. This computer package allows to
fit all the frequencies simultaneously in the magnitude domain.

\begin{figure}[]
\includegraphics[width=7.3cm]{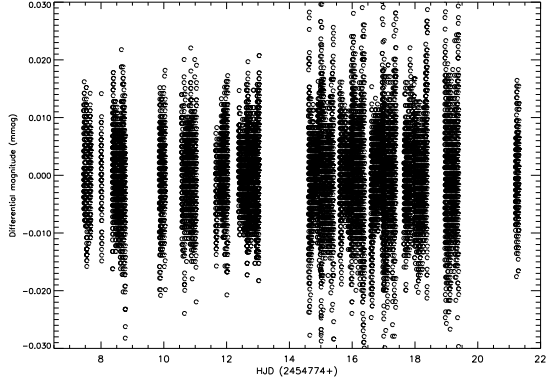}
\caption{Differential light curve V650 Tau $-$ Comparison 2.}
\label{fig:curves}
\end{figure}

\begin{figure}[!t]
 \centering
 \includegraphics[width=7.7cm]{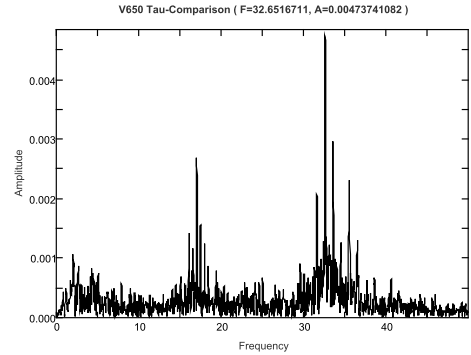}
 \caption{Amplitude spectrum derived from the light curve V650 Tau - Comparison 2. The amplitude is in mag and
 the frequencies in c/d.}
 \label{fig:spec}
\end{figure}

\medskip
The amplitude spectrum  of the differential light curve V650 Tauri -
Comparison 2  (see Fig~\ref{fig:curves}) is shown in
Figure~\ref{fig:spec}.  As can be seen, V650 Tauri presents
high-amplitude peaks distributed between 17 c/d and 35 c/d.

\medskip
The frequencies have been extracted by means of standard
prewhitening method. In order to decide which of the detected peaks
in the amplitude spectrum can be regarded as intrinsic to the star
we follow  Breger's criterion given by [7],  where it was shown that
the signal-to-noise ratio (in amplitude) should be at least  4 in
order to ensure that the extracted frequency is significant.

\medskip
The frequencies, amplitudes and phases  are listed in
Table~\ref{tab:frec}. Five significant frequencies have been
detected in V650 Tauri. Among these only the first two frequencies
(35.66 c/d and 17.04 c/d) are similar to that found by Kim \& Lee
(1996).
 A detailed analysis of these observations will be given in a forthcoming
paper [8].

\begin{table}[!t]\centering
  \setlength{\tabcolsep}{1.0\tabcolsep}
 \caption{Frequency peaks detected in the light curve V650 Tauri - Comparison 2. S/N
is the signal-to-noise ratio in amplitude after the prewhitening
process.} \label{tab:frec}
  \begin{tabular}{cccc}
\hline
Freq.&  A & $\varphi$/($2\pi$) & $S/N$  \\
(c/d)&(mmag)&&\\
\hline
       32.6565 &  4.43 &    0.39  &  13.4\\
       17.0356 &  2.65 &    0.09  &  10.0\\
       35.5970&   2.49 &    0.25  &   8.6\\
       31.6285 & 1.91 &   0.85  &  6.4\\
        34.6409&  1.41 &  0.04  &  4.5\\
 \hline
\end{tabular}
\end{table}

\begin{theacknowledgments}
This work has received financial support from the UNAM under grants
PAPIIT  IN108106 and IN114309. Special thanks are given to the
technical staff and night assistants of the San Pedro M\'artir,
Teide and Xing Long Observatories.  This research has made use of
the SIMBAD database operated at CDS, Strasbourg (France).

\end{theacknowledgments}



\bibliographystyle{aipprocl} 


\IfFileExists{\jobname.bbl}{}
 {\typeout{}
  \typeout{******************************************This is due to the worse windows function of out}
  \typeout{** Please run "bibtex \jobname" to optain}
  \typeout{** the bibliography and then re-run LaTeX}
  \typeout{** twice to fix the references!}
  \typeout{******************************************}
  \typeout{}
 }



\end{document}